\documentclass[12pt,a4,english]{extarticle}
\usepackage[utf8]{inputenc}
\usepackage{float}
\usepackage{units}
\usepackage{algorithm2e}

\usepackage{bm}
\usepackage{units}
\usepackage{mathtools}
\usepackage{amsmath}
\usepackage{amssymb}
\usepackage{cancel}
\usepackage[top=2.9cm, bottom=2cm, left=2.4cm, right=2.4cm]{geometry}
\usepackage{wasysym}
\usepackage{esint}

\usepackage[%
    unicode=true,
    bookmarks=true,
    colorlinks=true,
    linkcolor=black,
    citecolor=black,
    urlcolor=blue,
    filecolor=black,
    pdftitle={Implicit Large-Eddy Simulation of Two-Dimensional Kolmogorov Flow with Bottom Friction: Regimes, Cascades and Spectral Corrections},
    pdfauthor={Fedor Belolutskiy},
    pdfkeywords={two-dimensional turbulence, Kolmogorov flow, direct enstrophy cascade, inverse energy cascade, computational fluid dynamics, ILES, direct enstrophy cascade, inverse energy cascade, bottom friction},
    pdfproducer={LaTeX},
    pdfcreator={pdflatex},
    backref=true
]{hyperref}

\usepackage{graphicx}
\DeclareGraphicsExtensions{.png,.jpg,.pdf}

\makeatletter

\usepackage[english]{babel}
\usepackage{setspace}
\usepackage[T1]{fontenc}
\usepackage{newtxtext}
\usepackage{newtxmath}
\usepackage{fix-cm}
\usepackage{changepage}
\usepackage[autostyle, style=british]{csquotes}

\usepackage{titlesec}
\renewcommand{\thesection}{\arabic{section}}
\titleformat{\section}
  {\normalfont\centering} % Format: large + small caps + centered
  {\thesection.}                         % Label: section number
  {0.33em}
  {\MakeUppercase}                                    % Code before title
\titlespacing*{\section}
  {0pt}                                 % Left margin
  {4mm}                                 % Vertical space before (your \vspace*{4mm})
  {2mm}
\titleformat{name=\section,numberless}  % This targets \section*
  {\normalfont\centering}
  {}                                    % No section number
  {0pt}                                 % No separation since no number
  {\MakeUppercase}
\titlespacing*{\section}
  {0pt}{4mm}{2mm}% {4mm}
\renewcommand{\thesubsection}{\thesection.\arabic{subsection}}
\titleformat{\subsection}
  {\normalfont\centering}
  {\textit{\thesubsection.}}
  {0.33em}
  {\textit}
\titlespacing*{\subsection}
  {0pt}
  {1\baselineskip}
  {0pt}

\renewcommand{\thesubsubsection}{\thesubsection.\arabic{subsubsection}}

\titleformat{\subsubsection}
  {\normalfont\centering}                % same as subsection: normal font, centered
  {\textit{\thesubsubsection.}}          % label e.g. "2.1.1." in italic
  {0.33em}                               % separation between label and title
  {\textit}                              % title also in italic (as in subsection)
\titlespacing*{\subsubsection}
  {0pt}                                  % left margin
  {1\baselineskip}                       % space before (same as subsection)
  {0pt}                                  % space after

\usepackage{caption}
\makeatletter
\renewcommand{\fnum@figure}{\textbf{Fig.~\thefigure}}
\makeatother

\usepackage{indentfirst} % This package forces indentation of first paragraph

\makeatother

\makeatletter
\renewcommand{\maketitle}{%
  \begin{center}
    \textbf{\@title}%
  \end{center}
  \par
  \vspace{-\parskip}% optional: remove extra spacing if needed
}
\makeatother

\renewenvironment{abstract}{%
  \par\noindent\textbf{\abstractname---}\ignorespaces
}{%
  \par
}
\renewcommand{\abstractname}{Abstract}

\usepackage[backend=biber,
            style=numeric,
            sorting=none,       % order of citation (like unsrt)
            natbib=true,        % so \citep and \citet still work
            defernumbers=true,
            maxbibnames=99,
            giveninits=true]{biblatex}

\AtEveryBibitem{%
  \clearfield{url}%
  \clearfield{urldate}%
}

\DeclareFieldFormat{doi}{%
  \normalfont
  \ifhyperref
    {\ \href{https://doi.org/#1}{\textcolor{black}{https://doi.org/#1}}}%
    {\ https://doi.org/#1}%
  \adddot} % Add full stop at the end

\DeclareFieldFormat{labelnumberwidth}{#1\adddot}
\DeclareFieldFormat[article]{title}{\mkbibquote{#1}}

\DeclareFieldFormat{journaltitle}{#1}

\DeclareFieldFormat[article]{volume}{\textbf{#1}}

\DeclareFieldFormat[article]{number}{\mkbibparens{#1}}

\DeclareFieldFormat{pages}{#1}

\DeclareFieldFormat[article]{year}{\mkbibparens{#1}}

\DeclareFieldFormat[article,inproceedings,thesis]{title}{``#1\addcomma''}

\DeclareBibliographyDriver{article}{%
  \usebibmacro{bibindex}%
  \usebibmacro{begentry}%
  \usebibmacro{author/translator+others}%
  \setunit{\addcomma\space}%
  \usebibmacro{title}%
  \setunit{\space}%
  \printfield{journaltitle}%
  \setunit*{\addspace}%
  \printfield{volume}%
  \printfield{number}%
  \setunit{\addcomma\space}
  \printfield{pages}%
  \setunit*{\addspace}%
  \printfield{year}%
  \setunit{\adddot}%
  \printfield{doi}%
  \usebibmacro{finentry}%
}

\DeclareBibliographyDriver{book}{%
  \usebibmacro{bibindex}%
  \usebibmacro{begentry}%
  \normalfont\printnames{author}%
  \setunit{\normalfont\addcomma\space}\newblock
  \usebibmacro{title}% Title without quotes for books
  \setunit{\addspace}%
  \printtext{(}\printlist{publisher}%
  \setunit{\normalfont\addcomma\space}%
  \printlist{location}%
  \setunit{\normalfont\addcomma\space}%
  \printfield{year}%
  \printtext{)}%
  \usebibmacro{finentry}}

\DeclareBibliographyDriver{inproceedings}{%
  \usebibmacro{bibindex}%
  \usebibmacro{begentry}%
  \usebibmacro{author/translator+others}%
  \setunit{\addcomma\space}%
  \usebibmacro{title}%
  \setunit{\space}%
  \printtext{in\space}%
  \printfield[title]{booktitle}%
  \setunit{\addspace}%
  \printtext{\mkbibparens{%
    \printfield{location}%
    \setunit{\addcomma\space}%
    \printfield{year}%
  }}%
  \setunit{\addcomma\space}%
  \printtext{pp.\space}%
  \printfield{pages}%
  \setunit{\adddot}%
  \printfield{doi}%
  \usebibmacro{finentry}%
}

\DeclareFieldFormat[thesis]{title}{#1}

\DeclareFieldFormat[thesis]{type}{#1}

\renewbibmacro*{institution+location+date}{%
  \printtext[parens]{%
    \printlist{institution}%
    \setunit*{\addcomma\space}%
    \printlist{location}%
    \setunit*{\addcomma\space}%
    \usebibmacro{date}%
  }%
}

\AtEveryBibitem{%
  \ifentrytype{thesis}{%
  }{}%
}

\DeclareSourcemap{
  \maps[datatype=bibtex]{
    \map{
      \pertype{phdthesis}
      \step[fieldset=type, fieldvalue={Ph. D. Thesis}, final]
    }
    \map{
      \pertype{mastersthesis}
      \step[fieldset=type, fieldvalue={M. S. Thesis}, final]
    }
  }
}

\DeclareBibliographyDriver{thesis}{%
  \printnames{author}%
  \newunit\newblock
  \printfield{title}%
  \newunit\newblock
  \printfield{type}%
  \setunit{\addspace}%  % ! space
  \usebibmacro{institution+location+date}%
  \newunit\newblock
  \finentry}

\renewbibmacro{in:}{}

\usepackage{latexml}
\iflatexml
  
\else

\begin{document}
\vspace*{0mm}

% UDK 519.63.
% ICAD RAS Preprint. Dated: \today
{\footnotesize{\noindent This is a preprint of an article accepted for publication in {\textit{Mathematical Models and Computer Simulations}} \textbf{18}, Suppl. 1 (2026). © Pleiades Publishing, Ltd., 2026. The final version can be found at \href{https://pleiades.online}{https://pleiades.online}.}}

% \vspace*{0mm}
\title{Implicit Large-Eddy Simulation of Two-Dimensional Kolmogorov Flow with Bottom Friction: Regimes, Cascades and Spectral Corrections}
% \author{Fedor Belolutskiy (fedor1113-public@yandex.com), Svetlana Fortova (sfortova@mail.ru) \\
% KIAM RAS, Moscow, Russia}
% \date{\today}
\maketitle
\begin{center}
% \fontsize{11}{12.65}\selectfont
{\rm%\bfseries
Fedor Belolutskiy (\href{https://orcid.org/0000-0001-5807-3024}{\textcolor{black}{https://orcid.org/0000-0001-5807-3024}})$^{a,*}$ and Svetlana Fortova$^{a,**}$ 
% (\href{https://orcid.org/0000-0001-7444-9925}{\textcolor{black}{https://orcid.org/0000-0001-7444-9925}})$^{a,**}$
}\\
\vspace*{3mm}
\normalsize
\textit{$^{a}$KIAM RAS, Moscow, Russia} \\
% \textit{$^{b}$Skoltech, Moscow, Russia} \\
\textit{$^{*}$e--mail: \href{mailto:fedor1113-public@yandex.com}{\textcolor{black}{fedor1113-public@yandex.com}}} \\
\textit{$^{**}$e--mail: \href{mailto:sfortova@mail.ru}{\textcolor{black}{sfortova@mail.ru}}}
\medskip
\end{center}
\begin{abstract}
Low-Mach-number two-dimensional Kolmogorov flows with bottom friction are studied using implicit large-eddy simulation (ILES). The numerical modelling of the Euler equations is performed based on a fifth-order linear upwind finite-difference scheme for spatial discretisation, paired with an explicit fifth-order Runge--Kutta method for temporal integration. Flow regimes of a perfect gas with $ \gamma=1.4 $ and a speed of sound of the order of
$ \unit[100]{\text{m}/\text{s}} $ are considered for various bottom-friction coefficients $ \alpha $ and constant-in-time bidirectional forcing in a square periodic domain. We analyse the spectra of energy, enstrophy and their spectral fluxes for these flows depending on bottom friction. It is shown that the presence of friction significantly affects the spectra.
The associated spectral corrections are described and quantified.
\end{abstract}

\vspace*{2mm}

\vspace*{3mm}

\section{Introduction}

Academician~A.~N.~Kol\-mo\-go\-rov proposed a physical model that, to this day, serves as a testbed for two-dimensional turbulent flows sustained by periodic-in-space and constant-in-time external forcing \citep{Arnold:1960}.
Fluid motion forced in this way is referred to as Kolmogorov flow.
Such flows can be created in laboratory experiments by means of electromagnetic forces arising inside a thin layer of conducting fluid in a cell positioned above an array of magnets \citep{Bondarenko:1979,Obukhov:1983}.
% As later discovered by F.~V.~Dolzhanskii
As was later discovered by F.~V.~Dol\-zhan\-sky on the basis of the described laboratory experiments and other empirical data \citep{Bondarenko:1979,Obukhov:1983,Dolzhanskii:1990}, the relevant dissipative mechanisms in the Kolmogorov flow are not limited to internal molecular viscosity but also include bottom friction (drag), which must be accounted for in the Navier--Stokes equations via a linear momentum sink.
This flow type has received considerable attention over the past seven decades and continues to be studied analytically, numerically and experimentally \citep{Fylladitakis:2018}.

In particular, Kolmogorov flow of a viscous liquid under bidirectional forcing $ G \sin \left( k_f y \right) \boldsymbol{\hat{\imath}} -G \sin \left( k_f x \right) \boldsymbol{\hat{\jmath}} $ has recently been studied within the weak-compressibility approximation in the computational works \citep{Fortova:2026,Posudnevskaya:2026}.
As demonstrated therein, the governing parameters of Kolmogorov flow are the amplitude $G$ and wavenumber $k_f$ of the forcing ($ \ell_f = 2\pi/k_f $) as well as the bottom-friction coefficient $\alpha$.
For fixed forcing, under high enough friction, the flow is laminar and consists of a periodic system of counter-rotating vortices.
For intermediate values of the friction coefficient $ \alpha $, the flow becomes turbulent, and vortices of various scales and lifetimes start to form (chaotic regime).
Finally, with a low enough friction coefficient $ \alpha $, large coherent vortices with a clearly defined mean velocity profile appear (vortical regime).
These types of flow are driven by external forcing corresponding to high Reynolds and low Mach numbers.

At sufficiently weak bottom friction, the size of the vortices created by the energy-injecting force increases until they reach the size of the domain under consideration. This leads to energy accumulation at the corresponding scale.
The process is determined by the inverse energy cascade. Under periodic boundary conditions, the formation of a vortex dipole is observed \citep{Fortova:2026}.

In two-dimensional flows, the inverse energy cascade is accompanied by a direct enstrophy cascade.
The cascades are described by the incompressible Kraichnan--Batchelor theory \citep{Kraichnan:1967,Leith:1968,Batchelor:1969,Kraichnan:1971}. 
According to this theory, the dependence of the energy spectrum on the wavenumber in the inverse cascade corresponds to a power law with exponent \enquote{$ -5/3 $}: $ E\left( k \right) \propto k^{-5/3} $.
The spectrum of enstrophy in its direct cascade obeys $ Z \left( k \right) =k^2E\left(k\right) \propto k^{-1} \left( \ln \left( k/k_f \right) \right)^{-1/3} $.
The presence of bottom friction introduces corrections to both the slope of the spectrum in the inverse energy cascade \citep{Borue:1994,Smith:1994,Vallgren:2011,Alexakis:2018}
and the spectra in the inertial range of the direct enstrophy cascade \citep{Lilly:1972,Nam:2000,Valadao:2025,Ventrella:2026}.
The present study continues research in this area within the context of two-dimensional low-Mach-number Kolmogorov flow of a gas with bidirectional forcing.
This study uses the implicit large-eddy simulation (ILES) approach, where the flows are governed by the Euler equations instead of the viscous weakly compressible Navier--Stokes equations used in \citep{Fortova:2026,Posudnevskaya:2026}.
% What distinguishes the methodology of this study from \citep{Fortova:2026, Posudnevskaya:2026} is that the flows are governed by the Euler equations instead of the viscous weakly compressible Navier--Stokes equations.

To study the statistical and dynamical properties of two-dimensional turbulence in the presence of friction, we employ the concept of \enquote{rational} averaging for turbulent flows proposed by Academician~O.~M.~Belotserkovskii \citep{Belotserkovskii:1985,Belotserkovskii:2005turbulence}
% --- the implicit large-eddy simulation (ILES) approach
--- the aforementioned ILES approach
\citep{Grinstein:2007}. As O.~M.~Belotserkovskii noted \citep{Belotserkovskii:2005turbulence}:
% \begin{quote}
\enquote{There is a strong tendency for large-scale motions in turbulence to self-regulat[e,] resulting in independence of the large-scale structures [from] the details of the [viscous] dissipation mechanism [at small scales].}
% \end{quote}
In ILES, the scheme-induced hyperdiffusion
$ \boldsymbol{D}_{2n} = \nu_{2n}\left(\boldsymbol{U},\Delta x\right) \boldsymbol{\nabla}^{2n} + O\left(\Delta x^{2n}\right) $ serves as the dissipation mechanism instead of physical viscosity,
where $ \nu_{2n} $ is the kinematic hyperdiffusion coefficient of order $ 2n \in \mathbb{N} $.
This approach is actively used in studies of compressible two-dimensional turbulence \citep{Falkovich:2017,Puggioni:2020}, and, as reported in \citep{Kent:2011}, it is also used in atmospheric modelling. With regard to Kolmogorov flow specifically, we are aware of the paper \citep{Fortova:2018}, where the effects of unidirectional forcing were studied using ILES results.
These results once again emphasise that the onset of turbulence can occur in the absence of walls or viscosity.

In the present work, we propose a method based on an explicit fifth-order Runge--Kutta scheme for time integration \citep{Macdonald:2003,Henrick:2006} and a linear fifth-order upwind finite-difference scheme in space.
The method exhibits sixth-order hyperdiffusion $ \boldsymbol{D}_{6} \approx \nu_{6}\boldsymbol{\nabla}^{6} $ in its modified equation. We demonstrate that, although our model includes compressibility, its contribution to the energy balance is negligible for the considered low-Mach-number flows. The approach reproduces all regimes of Kolmogorov flow on a uniform $ 513\times513 $ grid with periodic boundary conditions. We obtain the dependence of the correction $\xi > 0$ to the spectral slope in the direct enstrophy cascade on the bottom-friction coefficient $ \alpha $, with $ Z \left( k \right) \propto k^{-1-\xi} \left( \ln \left( k/k_f \right) \right)^{-1/3} $. The energy spectra in the inverse cascade are also investigated for different values of $ \alpha $.

The paper is organised as follows. Section~\ref{sec:Mathematical-model-and-numerical-method} describes the mathematical model and the numerical method.
Section~\ref{sec:results} presents the results of numerical experiments and describes the flow regimes. The corresponding energy balance is discussed in Section~\ref{sec:balance}.
Section~\ref{sec:energy-spectra} analyses the energy and enstrophy spectra.
The conclusions are summarised in Section~\ref{sec:Conclusion}.

\section{Mathematical Model and Numerical Method}\label{sec:Mathematical-model-and-numerical-method}

\subsection{Mathematical Model}\label{subsec:Mathematical-model}

Consider a two-dimensional flow of a compressible inviscid continuum governed by the Euler equations with a bidirectional sinusoidal energy-injecting force and linear momentum losses due to bottom friction: 
\begin{equation}
\partial_{t}\overbrace{\begin{bmatrix}\rho\\
\rho u\\
\rho v\\
\rho \mathcal{E}
\end{bmatrix}}^{\boldsymbol{U}}+\partial_{x}\overbrace{\begin{bmatrix}\rho u\\
\rho u^{2}+p\\
\rho vu\\
\rho \mathcal{E}u+pu
\end{bmatrix}}^{\boldsymbol{F}_{x}}+\partial_{y}\overbrace{\begin{bmatrix}\rho v\\
\rho uv\\
\rho v^{2}+p\\
\rho \mathcal{E}v+pv
\end{bmatrix}}^{\boldsymbol{F}_{y}}=\begin{bmatrix}0\\
G\rho\sin\left(k_{f}y\right)\\
-G\rho\sin\left(k_{f}x\right)\\
G\rho\left(u\sin\left(k_{f}y\right)- v\sin\left(k_{f}x\right)\right)
\end{bmatrix}-\begin{bmatrix}0\\
\alpha\rho u\\
\alpha\rho v\\
\alpha\rho\left(u^{2}+v^{2}\right)
\end{bmatrix},\label{eq:euler-eqns}
\end{equation}
 where $\rho$ denotes density; $u,$ $v$ denote the components of the velocity vector;
$\mathcal{E}=\left(u^{2}+v^{2}\right)/2+e$ is the total specific energy; $e$ stands for the specific internal energy, while $\alpha$, $G$,
$k_{f}$ are the parameters of the problem. The equations are closed with the perfect-gas equation of state: 
\begin{equation}
e=\frac{p}{\left(\gamma-1\right)\rho}.\label{eq:perfect-gas-EoS}
\end{equation}
We fix the parameters of the equation of state and the external force such that the energy-injection wavenumber $k_{f}$ is much smaller than the wavenumber corresponding to the characteristic scale of hyperdiffusion, $k_{h\nu}$ ($k_{f}\ll k_{h\nu}$):
\begin{equation}
\gamma=1.4;\qquad G=\unit[0.15]{\text{m}/\text{s}^{2}};\qquad k_{f}=\unit[5]{\text{m}^{-1}}.
\end{equation}

The bottom-friction coefficient $ \alpha $ takes the following values: $\unit[1]{\text{s}^{-1}}$,
$\unit[0.3]{\text{s}^{-1}}$, $\unit[0.1]{\text{s}^{-1}}$, $\unit[0.01]{\text{s}^{-1}}$,
$\unit[0.0001]{\text{s}^{-1}}$.
The listed values correspond, in order, to the laminar, laminar--chaotic, chaotic, chaotic--vortical and vortical regimes for a two-dimensional Kolmogorov flow of a viscous weakly compressible fluid, as reported in \citep{Fortova:2026}.

Let us specify the initial and boundary conditions for the numerical experiments conducted in the present work.
The medium is initially at rest ($u=v=0$) with uniform pressure
$p=\unit[10^{5}]{\text{Pa}}$ and density $\rho=\rho_{0}=\unit[14]{\text{kg}/\text{m}^{3}}$
(to match the speed of sound from \citep{Fortova:2026}, which is equal to $c_{s}=\sqrt{\gamma p/\rho}=\unit[100]{\text{m}/\text{s}}$). The boundary conditions are periodic in both directions within a square computational domain $\Omega=\left[0\mathop{..}\unit[2\pi]{\text{m}}\right]\times\left[0\mathop{..}\unit[2\pi]{\text{m}}\right]$.

\subsection{Numerical Method}\label{subsec:Numerical-Method}

We use ILES based on a conservative finite-difference scheme with flux-vector splitting (FVS) to numerically integrate \eqref{eq:euler-eqns}--\eqref{eq:perfect-gas-EoS}.
% In accordance with the standard method of lines (MoL), we first semi-discretise the system in space and then integrate the resulting ordinary differential equation at each grid point.
As a spatial discretisation, we utilise a linear component-wise upwind scheme with fifth-order approximation and global Lax--Friedrichs flux splitting. As a temporal integrator, we make use of a suitably high-order explicit Runge--Kutta method. The flows considered here have low Mach numbers, so a high-order linear interpolation for the split fluxes essentially does not create unphysical oscillations.

Global Lax--Friedrichs flux splitting is performed dimension by dimension at every point of the grid with coordinates $ x_{i}=i\Delta x $ and $ y_{j}=j\Delta y $,
$ i,j=0,\dots,N $ ($ N=512 $) at each time step $ t^{k} $:
\begin{equation}
\hat{\boldsymbol{f}}^{\left(\pm\right),k}_{i,j}=\frac{1}{2}\left(\boldsymbol{F}_{x}\left(\boldsymbol{U}_{i,j}\right)\pm\mathop{\underset{\Omega}{\max}}\left|\lambda\right|\boldsymbol{U}_{i,j}\right)
\end{equation}
 for the $ X $ axis and 
\begin{equation}
\hat{\boldsymbol{f}}^{\left(\pm\right),k}_{i,j}=\frac{1}{2}\left(\boldsymbol{F}_{y}\left(\boldsymbol{U}_{i,j}\right)\pm\mathop{\underset{\Omega}{\max}}\left|\lambda\right|\boldsymbol{U}_{i,j}\right)
\end{equation}
 for the $ Y $ axis, where $\mathop{\underset{\Omega}{\max}}\left|\lambda\right|=\mathop{\underset{i,j=0,\dots,N}{\max}}\left\{ \left|u^{k}_{i,j}\right|+{c_{s}}^{k}_{i,j},\left|v^{k}_{i,j}\right|+{c_{s}}^{k}_{i,j}\right\} $,
while $\boldsymbol{U}_{i,j}=\boldsymbol{U}\left(x_{i},y_{j}\right)$.

The linear upwind scheme interpolates these split fluxes at the interfaces $ i+1/2 $, $ j+1/2 $.
Let us describe the interpolation for fluxes through interfaces $i+1/2$ directed along the $X$ axis  and the negative $X$ axis. The following formula is used for non-negative fluxes along the $X$ axis: 
\begin{equation}
\hat{\boldsymbol{f}}^{\left(+\right),k}_{i+1/2,j}=\frac{2\boldsymbol{\hat{f}}^{\left(+\right),k}_{i-2,j}-13\boldsymbol{\hat{f}}^{\left(+\right),k}_{i-1,j}+47\boldsymbol{\hat{f}}^{\left(+\right),k}_{i,j}+27\boldsymbol{\hat{f}}^{\left(+\right),k}_{i+1,j}-3\boldsymbol{\hat{f}}^{\left(+\right),k}_{i+2,j}}{60},\label{eq:lup5}
\end{equation}
 such that 
\begin{equation}
\partial_{x}\left[\hat{\boldsymbol{f}}^{\left(+\right),k}\right]_{i,j}=D^{+}_{\Delta x}\left[\hat{\boldsymbol{f}}^{\left(+\right),k}\right]_{i,j}+O\left(\Delta x^{5}\right),
\end{equation}
 where 
\begin{equation}
D^{+}_{\Delta x}\left[\hat{\boldsymbol{f}}^{\left(+\right),k}\right]_{i,j}=\frac{1}{\Delta x}\left(\hat{\boldsymbol{f}}^{\left(+\right),k}_{i+1/2,j}-\hat{\boldsymbol{f}}^{\left(+\right),k}_{i-1/2,j}\right),
\end{equation}

A similar formula is used to obtain the upwind approximation for fluxes running in the negative $X$-direction: 
\begin{equation}
\hat{\boldsymbol{f}}^{\left(-\right),k}_{i+1/2,j}=\frac{2\boldsymbol{\hat{f}}^{\left(-\right),k}_{i+3,j}-13\boldsymbol{\hat{f}}^{\left(-\right),k}_{i+2,j}+47\boldsymbol{\hat{f}}^{\left(-\right),k}_{i+1,j}+27\boldsymbol{\hat{f}}^{\left(-\right),k}_{i,j}-3\boldsymbol{\hat{f}}^{\left(-\right),k}_{i-1,j}}{60}.
\end{equation}
\enlargethispage{2\baselineskip}
An analogous approach is used for fluxes on the $Y$ axis.
Further, all the split fluxes in all directions are summed as 
\begin{equation}
\boldsymbol{S}^{\Delta x,k}_{i,j}=-\frac{1}{\Delta x}\left(\hat{\boldsymbol{f}}^{\left(+\right),k}_{i+1/2,j}+\hat{\boldsymbol{f}}^{\left(-\right),k}_{i+1/2,j}-\hat{\boldsymbol{f}}^{\left(+\right),k}_{i-1/2,j}-\hat{\boldsymbol{f}}^{\left(-\right),k}_{i-1/2,j}+\hat{\boldsymbol{f}}^{\left(+\right),k}_{i,j+1/2}+\hat{\boldsymbol{f}}^{\left(-\right),k}_{i,j+1/2}-\hat{\boldsymbol{f}}^{\left(+\right),k}_{i,j-1/2}-\hat{\boldsymbol{f}}^{\left(-\right),k}_{i,j-1/2}\right),
\end{equation}
 and the resulting ordinary differential equation at time $t^{k}=t^{k-1}+\Delta \tau $
\begin{equation}
\frac{d}{dt}\boldsymbol{U}^{k}_{i,j}\approx\boldsymbol{S}^{\Delta x,k}_{i,j}\label{eq:mol}
\end{equation}
is integrated by C.~B.~Mac\-do\-nald's explicit six-stage fifth-order Runge--Kutta method \citep{Macdonald:2003,Henrick:2006}. 
All appropriate pointwise source and sink terms on the right-hand side of \eqref{eq:euler-eqns} at $ \left(x_{i},y_{j}\right) $ are also added to $\boldsymbol{S}^{\Delta x,k}_{i,j}$ in equation \eqref{eq:mol} before integration.

For each $t^{k}$, the time step of the scheme is determined via
$\Delta \tau=\text{CFL}\Delta x/\mathop{\underset{\Omega}{\max}}\left|\lambda\right|$, subject to the Cou\-rant--Friedrichs--Lewy condition.
The properties of the high-order Runge--Kutta method allow the Courant number to be set to $\text{CFL}=1.2$, similarly to what is done in \citep{Henrick:2006}.
Flow snapshots are saved every $ \Delta t = \unit[0.25]{\text{s}} $.

The numerical scheme described here can be used in ILES since it is dissipative: it possesses numerical hyperdiffusion.
Let us estimate the characteristic scale of this hyperdiffusion.
Fixing the energy injection (forcing) at wavenumber $k_{f}=\unit[5]{\text{m}^{-1}}$ with a root-mean-square flow velocity
not greater than $10$--$\unit[20]{\text{m}/\text{s}}$ and a grid of $513\times513$ points over the computational domain $ \Omega $ allows us to estimate the hyperdiffusion scale as follows.
The first differential approximation (the modified equation) of the numerical scheme for a linearised problem is 
\begin{equation}
\partial_{t}\boldsymbol{U}+\partial_{x}\boldsymbol{F}_{x}\left(\boldsymbol{U}\right)+\partial_{y}\boldsymbol{F}_{y}\left(\boldsymbol{U}\right)\approx\nu_{6}\left(\boldsymbol{U},\Delta x,\text{CFL}\right)\left(\partial^{6}_{x}+\partial^{6}_{y}\right)\boldsymbol{U},
\end{equation}
 where the kinematic sixth-order hyperdiffusion coefficient for
$\text{CFL}=1.2$ can be estimated as
\begin{equation}
\nu_{6}=\nu_{6}\left(\boldsymbol{U},\Delta x\right)
% \sim\left(\frac{1}{60}-\frac{689\cdot\left(\text{CFL}\right)^{5}}{450\,000}\right)\cdot\Delta x^{5}\cdot\mathop{\underset{\Omega}{\max}}\left|\lambda\right|
\sim0.013\cdot\Delta x^{5}\cdot\mathop{\underset{\Omega}{\max}}\left|\lambda\right|\sim\unit[3.6\cdot{10}^{-10}]{\text{m}^{6}/\text{s}}
\end{equation}
 with $ \lambda\sim c_{s}\sim\unit[100]{\text{m}/\text{s}} $ (which corresponds to the flows under consideration). Then an estimate based on dimensional analysis for the characteristic hyperdiffusion wavenumber is $ k_{h\nu} \sim \left(\nu_{6}/\eta^{1/3}_{f}\right)^{-1/6}\sim $
$ 36 $--$ \unit[47]{\text{m}^{-1}} $, where $ \eta_{f}=k^{2}_{f}\varepsilon_{f} $
is the rate of enstrophy injection.\footnote{Here $ \varepsilon_{f}=\left({4\pi^{2}}\right)^{-1}\iintop_{\Omega}G\left(u\sin\left(k_f y\right)- v\sin\left(k_f x\right)\right)dxdy $ denotes the rate of specific-energy injection due to the forcing.} The injection wavenumber $ k_{f} $ is much less than the characteristic hyperdiffusion wavenumber $ k_{h\nu} $, which allows for a direct enstrophy cascade to develop between $ k_{f} $ and $ k_{h\nu} $.

% \section{Results}\label{sec:results}
\section{Regimes of Kolmogorov Flow}\label{sec:results}

Let us consider the regimes of two-dimensional Kolmogorov flow in a square computational domain with periodic boundary conditions.
The flow dynamics depends on the external forcing and bottom friction. We examine the influence of bottom friction on the formation of various flow regimes for fixed forcing parameters.
Recall that in \citep{Fortova:2026}, numerical simulations of the weakly compressible Navier--Stokes equations for Kolmogorov flow yielded laminar, chaotic and vortical regimes with bottom-friction coefficients $\alpha$ taking the values $\unit[1]{\text{s}^{-1}}$, $\unit[0.3]{\text{s}^{-1}}$, $\unit[0.1]{\text{s}^{-1}}$, $\unit[0.01]{\text{s}^{-1}}$, $\unit[0.0001]{\text{s}^{-1}}$ at a forcing amplitude of $ G=\unit[0.15]{\text{m}/\text{s}^{2}} $. 
We show that at the same values of $ G $ and $ \alpha $, analogous Kolmogorov flow regimes can be obtained by performing simulations of the Euler equations using an implicit large-eddy simulation approach.

% \subsection{Flow regimes}\label{subsec:dynamical-regimes}

\begin{figure}[H]
\begin{centering}
\includegraphics[width=1\textwidth]{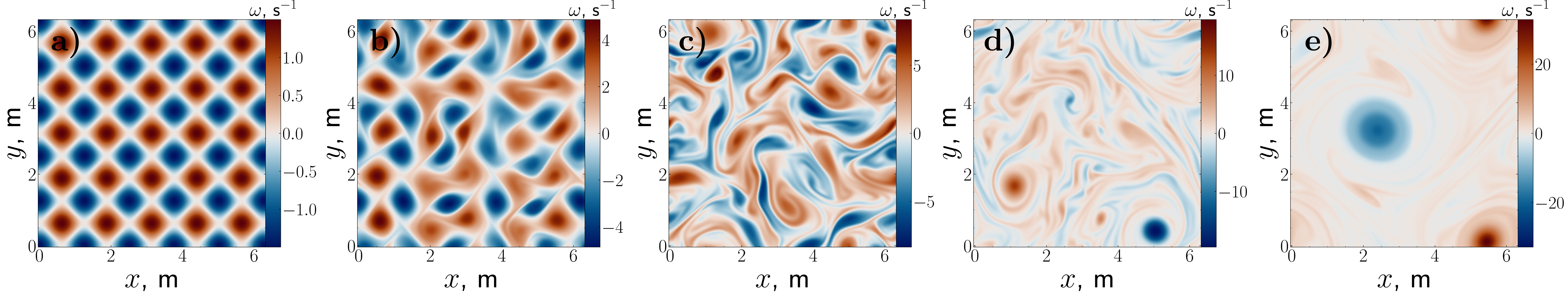}
\par\end{centering}
\caption{Instantaneous vorticity fields for the developed (a) laminar ($\alpha=\unit[1]{\text{s}^{-1}}$),
(b) laminar--chaotic ($\alpha=\unit[0.3]{\text{s}^{-1}}$),
(c) chaotic ($\alpha=\unit[0.1]{\text{s}^{-1}}$), (d) chaotic--vortical
($\alpha=\unit[0.01]{\text{s}^{-1}}$) and (e) vortical ($\alpha=\unit[0.0001]{\text{s}^{-1}}$)
regimes of Kolmogorov flow}\label{fig:vorticity}
\end{figure}

We turn our attention to various saturated flow regimes, all of which have reached a statistically steady state. The regimes are illustrated by their instantaneous vorticity fields $ \omega=\partial_{x}v-\partial_{y}u $.
Fig.~\ref{fig:vorticity} shows that as the friction coefficient $\alpha$ decreases, the flow regime changes from laminar ($\alpha=\unit[1]{\text{s}^{-1}}$) to chaotic ($\alpha=\unit[0.1]{\text{s}^{-1}}$) and then to vortical ($\alpha=\unit[0.0001]{\text{s}^{-1}}$), as in \citep{Fortova:2026}.
At a sufficiently large friction coefficient, the vorticity is driven linearly by the external forcing and forms a time-independent system of counter-rotating vortices (Fig.~\ref{fig:vorticity}a). 
As friction decreases, however, the flow loses stability and becomes turbulent, with the formation of a large number of vortical structures (eddies) of different scales (Fig.~\ref{fig:vorticity}c).
Finally, at a sufficiently low friction coefficient, an inverse energy cascade develops, creating large-scale coherent structures: a vortex dipole (Fig.~\ref{fig:vorticity}e).
The size of each vortex in the pair is about half the size of the computational domain.
Transitional regimes arise between the main flow regimes: laminar--chaotic ($\alpha=\unit[0.3]{\text{s}^{-1}}$, Fig.~\ref{fig:vorticity}b) and chaotic--vortical ($\alpha=\unit[0.01]{\text{s}^{-1}}$, Fig.~\ref{fig:vorticity}d). 
Note that the magnitude of vorticity increases with decreasing friction coefficient $\alpha$.

\section{Balance of Kinetic Energy in a Fully Developed Flow}\label{sec:balance}
% \subsection{Balance of kinetic energy in a fully developed flow}\label{subsec:energy-balance}
% \enlargethispage{2\baselineskip}
\enlargethispage{\baselineskip}

Since we are interested in fully developed flows, which have already reached a statistically steady state, we expect that the integral kinetic energy of the flow is also steady on average, with positive and negative fluctuations of comparable magnitude that cancel out.
The kinetic-energy balance for a compressible fluid in a periodic domain $\Omega$ is determined by the following equation, taking hyperdiffusion into account: 
\begin{equation}
\frac{d\varepsilon}{dt}=P_{f}+P_{pd}-D_{f}-D_{h\nu},\label{eq:kinetic_energy_balance}
\end{equation}
 where 
\begin{equation}
\varepsilon=\frac{1}{2}\iintop_{\Omega}\rho\left(u^{2}+v^{2}\right)dxdy\label{eq:kin-en}
\end{equation}
 is the integral kinetic energy of the system;
\pagebreak[4]
\begin{equation}
P_{f}=\iintop_{\Omega}\left(\rho uG\sin\left(k_f y\right)-\rho vG\sin\left(k_f x\right)\right)dxdy
\end{equation}
 is the integral power of the external force (energy injection rate);
\begin{equation}
P_{pd}=-\iintop_{\Omega}\left(u\cdot\boldsymbol{\nabla}p\right)dxdy=\iintop_{\Omega}p\left(\boldsymbol{\nabla}\cdot\boldsymbol{u}\right)dxdy
\end{equation}
 is the integral power due to pressure dilatation determined by the compressibility of the medium;
\begin{equation}
D_{f}=\alpha\iintop_{\Omega}\rho\left(u^{2}+v^{2}\right)dxdy=2\alpha\varepsilon
\end{equation}
 is the integral power of the frictional forces (energy dissipation rate due to friction);
\begin{equation}
D_{h\nu}\sim\nu_{6}\iintop_{\Omega}\rho\left|\boldsymbol{\nabla}^{3}\boldsymbol{u}\right|^{2}dxdy
\end{equation}
 is an estimate of the integral power of hyperdiffusive forces inherent in the numerical scheme (energy dissipation rate due to hyperdiffusion).
The hyperdiffusion coefficient $\nu_{6}$ depends on
$\rho\boldsymbol{u}$, $c_{s}$ and grid resolution.

\begin{figure}[H]
\begin{centering}
\includegraphics[width=0.59\textwidth]{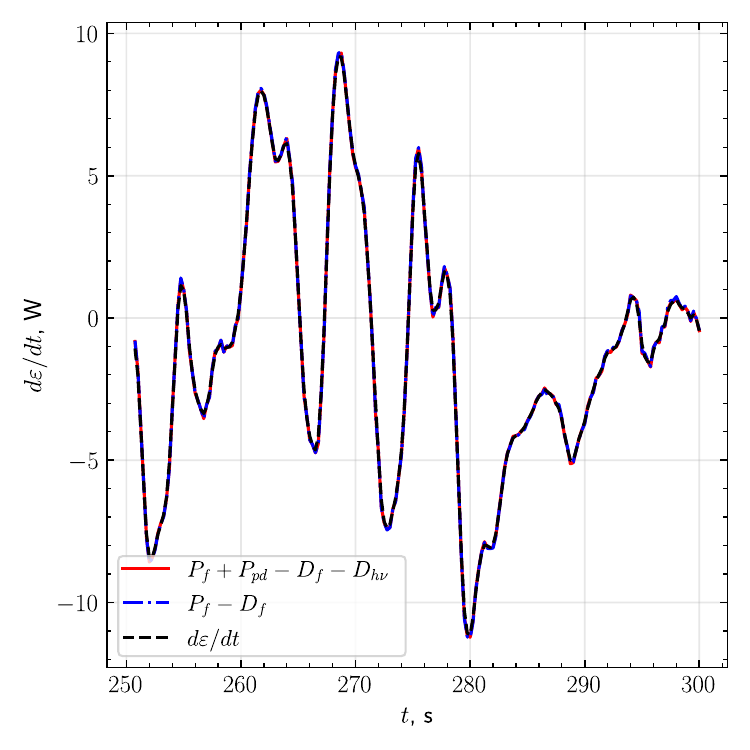}
\par\end{centering}
\caption{Energy balance for a developed vortical regime of the Kolmogorov flow with bottom friction $\alpha=\unit[0.0001]{\text{s}^{-1}}$}\label{fig:energy-budget}
\end{figure}

In the incompressible inviscid approximation, the balance reduces to
\begin{equation}
\frac{d\varepsilon}{dt}\approx P_{f}-D_{f},\label{eq:inviscid_kinetic_energy_balance}
\end{equation}
 which works well for the regimes considered here and in \citep{Posudnevskaya:2026}.

The kinetic-energy balance computed via \eqref{eq:kinetic_energy_balance} for a developed vortical regime of Kolmogorov flow with a low bottom-friction coefficient, $ \alpha=\unit[{10}^{-4}]{\text{s}^{-1}} $, is depicted in Fig.~\ref{fig:energy-budget}.
A two-point boxcar moving average is applied to all the time series in the figure.
We take $\nu_{6} = \unit[3.6\cdot{10}^{-10}]{\text{m}^{6}/\text{s}} $ to estimate $D_{h\nu}$. The solid red line corresponds to the right-hand side of \eqref{eq:kinetic_energy_balance}.
The dashed black line corresponds to the left-hand side, where $d\varepsilon/dt\approx\left(\varepsilon\left(t+\Delta t\right)-\varepsilon\left(t-\Delta t\right)\right)/\left(2\Delta t\right)$.
The blue dash-dot line is the right-hand side of the approximate formula
\eqref{eq:inviscid_kinetic_energy_balance}. These three lines are in close agreement.

As can be deduced from this plot,
the contributions of hyperdiffusion and compressibility to the energy balance are minimal.
It is evident that the relative contribution of friction can only increase for larger values of $\alpha$, with the relative contribution of hyperdiffusion decreasing accordingly.
This is confirmed by our simulations. 
Hence, energy dissipation is mainly determined by friction for the present problem configurations, in agreement with the numerical results for analogous problems in the viscous model of \citep{Posudnevskaya:2026}.

Let us also make a quantitative estimate of the contributions of hyperdiffusion and compressibility to the kinetic-energy balance for the vortical flow regime with $\alpha=\unit[{10}^{-4}]{\text{s}^{-1}}$.
The hyperdiffusion contribution $ D_{h\nu} $ is two orders of magnitude smaller than the friction contribution $ D_{f} $ and the energy injection contribution $ P_{f} $, while the compressibility contribution $P_{pd}$ is three orders of magnitude smaller than $P_{f}$ and $D_{f}$.
This means that the flows under consideration are effectively incompressible, and the Kraichnan--Batchelor theory can be applied to them.

\section{Energy and Enstrophy Spectra}\label{sec:energy-spectra}
% \subsection{Energy and enstrophy spectra}\label{subsec:energy-spectra}

We construct the time-averaged energy and enstrophy spectra sampled at every $ \Delta t = \unit[0.25]{\text{s}} $ for $t\in\left[\unit[200]{\text{s}}\mathop{..}\unit[300]{\text{s}}\right]$, once the flow is fully developed.

The fast Fourier transform (FFT) $\mathcal{F}$ is applied to the velocity components $u$, $v$ to extract the specific-kinetic-energy spectrum for a given time: 
\begin{equation}
E\left(k\right)=\frac{1}{2}\sum_{k-\frac{1}{2}\leq\left|\boldsymbol{k}\right|<k+\frac{1}{2}}\left(\left|\hat{u}\left( \boldsymbol{k} \right)\right|^{2}+\left|\hat{v}\left( \boldsymbol{k} \right)\right|^{2}\right),
\end{equation}
 where $\hat{u}\left( \boldsymbol{k} \right)=\mathcal{F}\left\{ u\right\} \left(\boldsymbol{k}\right)/N^{2}$,
$\hat{v}\left( \boldsymbol{k} \right)=\mathcal{F}\left\{ v\right\} \left(\boldsymbol{k}\right)/N^{2}$,
while $\boldsymbol{k}=\left(k_{x},k_{y}\right)$ are the wave vectors with integer wavenumbers, indices
$ k=1,\dots,N/2 $.
The instantaneous enstrophy spectrum $Z\left(k\right)$ as a function of wavenumber $k$ is computed similarly, where the integral enstrophy is defined as
\begin{equation}
Z=\frac{1}{2}\iintop_{\Omega}\omega^{2}dxdy=\frac{1}{2}\iintop_{\Omega}\left(\partial_{x}v-\partial_{y}u\right)^{2}dxdy.
\end{equation}

\subsection{Inverse Energy Cascade}\label{subsec:inverse_energy_cascade}
% \subsubsection{Inverse energy cascade}

\begin{figure}[H]
\begin{centering}
\includegraphics[width=1\textwidth]{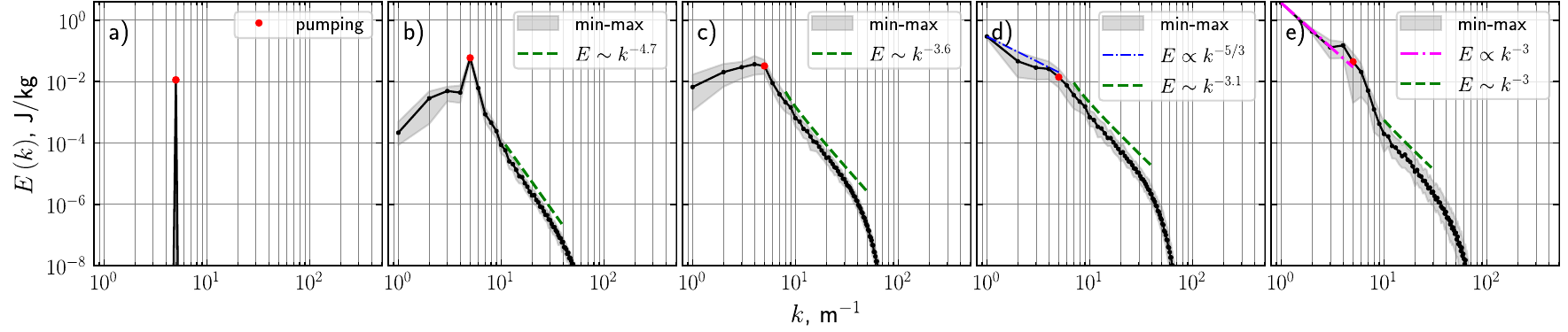}
\par\end{centering}
\caption{Averaged specific-kinetic-energy $E$ spectra
for the developed (a) laminar ($\alpha=\unit[1]{\text{s}^{-1}}$), (b) laminar--chaotic
($\alpha=\unit[0.3]{\text{s}^{-1}}$), (c) chaotic ($\alpha=\unit[0.1]{\text{s}^{-1}}$),
(d) chaotic--vortical ($\alpha=\unit[0.01]{\text{s}^{-1}}$) and
(e) vortical ($\alpha=\unit[0.0001]{\text{s}^{-1}}$) flow regimes
% ($k$ is measured in $\text{m}^{-1}$, while $E\left(k\right)$ is measured in $\unit{\text{J}/\text{kg}}=\unit{\text{m}^{2}/\text{s}^{2}}$)
}\label{fig:energy-spectra}
\end{figure}

Energy spectra corresponding to various bottom-friction coefficients are displayed in Fig.~\ref{fig:energy-spectra}.
Here and below, the variation of the spectra over the averaging time segment is indicated in grey.
The red dot marks the spectral value at the wavenumber $k_{f}=\unit[5]{\text{m}}^{-1}$,
where the forcing injects energy into the system.
We now describe how the formation of the inverse energy cascade depends on the bottom-friction coefficient $ \alpha $.
The kinetic energy is maximal at the injection wavenumber $k_{f}$ for large values of $ \alpha > \unit[0.1]{\text{s}^{-1}} $.
When the coefficient is lowered from $\alpha=\unit[0.1]{\text{s}^{-1}}$ to $\alpha=\unit[0.0001]{\text{s}^{-1}}$,
an inverse energy cascade develops, 
manifested as a change in the slope of the low-wavenumber part of the spectrum.
This change attests to the condensation of energy at large scales.

According to the Kraichnan--Batchelor theory, this cascade for an incompressible friction-free medium obeys the \enquote{$ -5/3 $} law \citep{Kraichnan:1967,Leith:1968,Batchelor:1969,Kraichnan:1971}.
If bottom friction is present, the power law for the inverse energy cascade can acquire a correction of the form $ E\left(k\right)\propto k^{-5/3-\mu} $
\citep{Vallgren:2011}.
For a developed vortical regime with $ \alpha=\unit[{10}^{-4}]{\text{s}^{-1}} $ in Fig.~\ref{fig:energy-spectra}e, the inverse energy cascade closely follows $ E\left(k\right)\propto k^{-3} $,
in agreement with \citep{Smith:1994,Vallgren:2011}.
% This is consistent with the $ k^{-3} $ scaling predicted for condensates in finite domains with bottom friction \citep{Smith:1994} and with the steepening of the inverse cascade observed in strongly vortical regimes \citep{Vallgren:2011}.

\begin{figure}[H]
\begin{centering}
\includegraphics[width=1\textwidth]{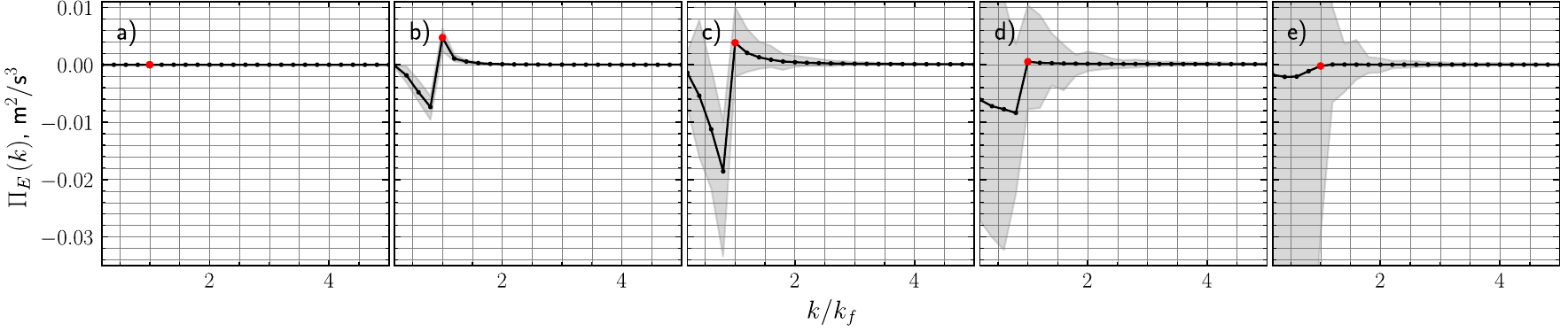}
\par\end{centering}
\caption{Average energy spectral fluxes $\Pi_{E}$ for
the developed (a) laminar ($\alpha=\unit[1]{\text{s}^{-1}}$), (b) laminar--chaotic
($\alpha=\unit[0.3]{\text{s}^{-1}}$), (c) chaotic ($\alpha=\unit[0.1]{\text{s}^{-1}}$),
(d) chaotic--vortical ($\alpha=\unit[0.01]{\text{s}^{-1}}$) and
(e) vortical ($\alpha=\unit[0.0001]{\text{s}^{-1}}$) flow regimes
% ($\Pi_{E}\left(k\right)$ is measured in $\text{m}^{2}/\text{s}^{3}$)
}\label{fig:energy-flux-spectra}
\end{figure}

The development of the inverse energy cascade can also be traced by means of spectral fluxes of energy.
As explained in \citep{Alexakis:2018},
the flux is positive when energy is transferred to smaller scales (direct cascade) and negative when energy is transferred back to larger scales (inverse cascade).
The spectral flux $ \Pi_E $ of specific kinetic energy for incompressible turbulence in a periodic domain is, following \citep{Alexakis:2018}, equal to:
\begin{equation}
\Pi_{E}\left(k\right)=\frac{1}{N^{2}}\sum_{k^{\prime}<k}\sum_{k^{\prime}-\frac{1}{2}\leq\left|\boldsymbol{k}\right|<k^{\prime}+\frac{1}{2}}\mathop{\Re\text{e}}\left[\overline{\mathcal{F}\left\{ u\right\} \left(\boldsymbol{k}\right)}\cdot\mathcal{F}\left\{ \left(\boldsymbol{u}\cdot\boldsymbol{\nabla}\right)u\right\} \left(\boldsymbol{k}\right)+\overline{\mathcal{F}\left\{ v\right\} \left(\boldsymbol{k}\right)}\cdot\mathcal{F}\left\{ \left(\boldsymbol{u}\cdot\boldsymbol{\nabla}\right)v\right\} \left(\boldsymbol{k}\right)\right],
\end{equation}
 where the overline denotes the complex conjugate,
and $\mathop{\Re\text{e}}$ is the real part of the complex number.
Fig.~\ref{fig:energy-flux-spectra} shows that the flux is non-positive for $ k < k_{f} $.
This means that energy is transferred from the injection scale to larger scales.
Thus an inverse energy cascade is established. 
Since the energy flux nearly vanishes for wavenumbers $ k > k_{f} $, no strong energy cascade is present there.

\subsection{Direct Enstrophy Cascade}\label{subsec:direct_enstrophy_cascade}
% \subsubsection{Direct enstrophy cascade}

\begin{figure}[H]
\begin{centering}
\includegraphics[width=1\textwidth]{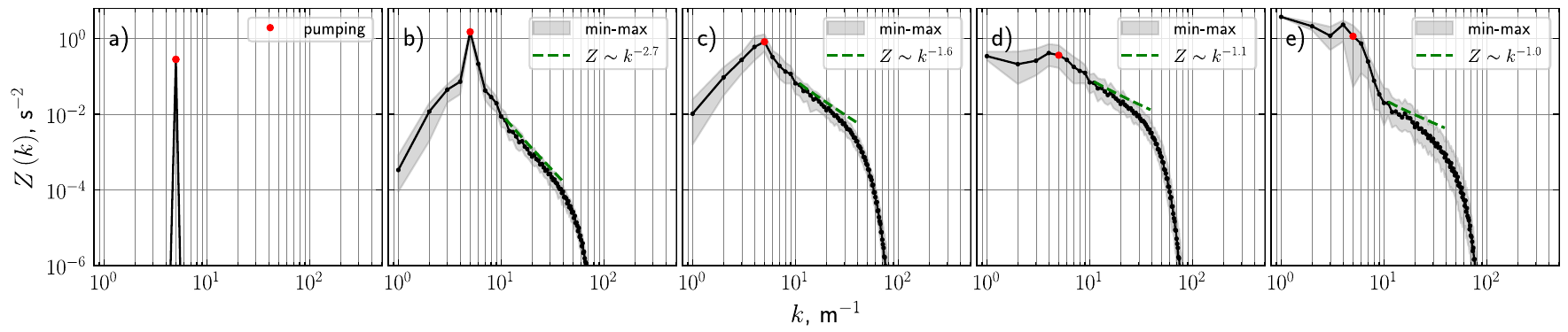}
\par\end{centering}
\caption{Averaged enstrophy $Z$ spectra for the developed (a) laminar ($\alpha=\unit[1]{\text{s}^{-1}}$), (b) laminar--chaotic
($\alpha=\unit[0.3]{\text{s}^{-1}}$), (c) chaotic ($\alpha=\unit[0.1]{\text{s}^{-1}}$),
(d) chaotic--vortical ($\alpha=\unit[0.01]{\text{s}^{-1}}$) and
(e) vortical ($\alpha=\unit[0.0001]{\text{s}^{-1}}$) flow regimes
% ($k$ is measured in $\text{m}^{-1}$, while $Z\left(k\right)$ is measured in $\text{s}^{-2}$)
}\label{fig:enstrophy-spectra}
\end{figure}

It has been established that the inverse energy cascade is accompanied by the direct enstrophy cascade in two-dimensional turbulence.
The enstrophy spectra are presented in Fig.~\ref{fig:enstrophy-spectra}.
The maximum of enstrophy is located near the injection wavenumber
$k_{f}$ in all flow regimes.
We now describe how the direct enstrophy cascade depends on the bottom-friction coefficient $ \alpha $.
When the friction coefficient is lowered from $ \alpha=\unit[1]{\text{s}^{-1}} $ to $ \alpha=\unit[0.0001]{\text{s}^{-1}} $, a direct enstrophy cascade appears,
manifested as a change in the slope of a part of the spectrum in the range of high wavenumbers.

According to the Kraichnan--Batchelor theory, enstrophy in this cascade obeys the following law \citep{Kraichnan:1967,Leith:1968,Batchelor:1969,Kraichnan:1971}: $ Z\left(k\right)\propto k^{-1}\left(\ln\left(k/k_{f}\right)\right)^{-1/3} $.

\begin{figure}[H]
\begin{centering}
\includegraphics[width=1\textwidth]{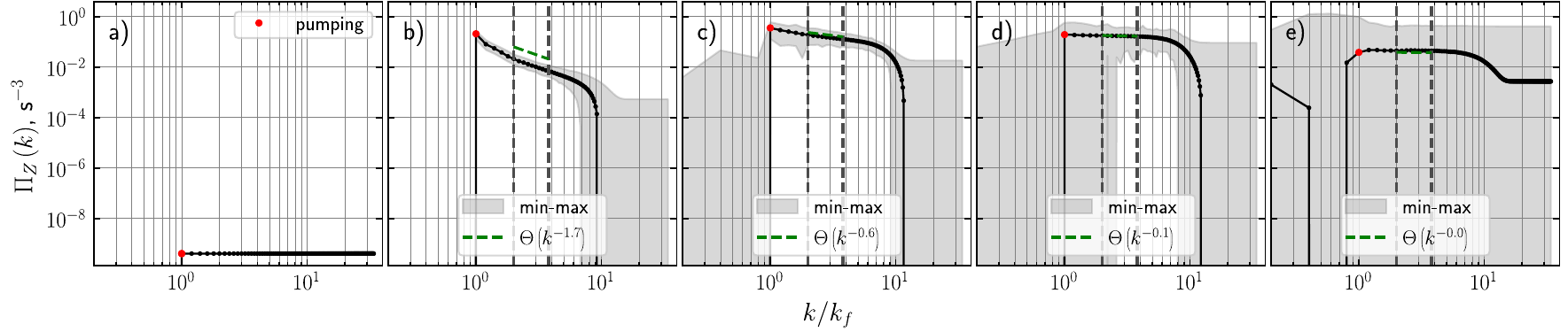}
\par\end{centering}
\caption{Average enstrophy spectral fluxes $\Pi_{Z}$ for the developed (a) laminar ($\alpha=\unit[1]{\text{s}^{-1}}$), (b) laminar--chaotic
($\alpha=\unit[0.3]{\text{s}^{-1}}$), (c) chaotic ($\alpha=\unit[0.1]{\text{s}^{-1}}$),
(d) chaotic--vortical ($\alpha=\unit[0.01]{\text{s}^{-1}}$) and
(e) vortical ($\alpha=\unit[0.0001]{\text{s}^{-1}}$) flow regimes
%($\Pi_{Z}\left(k\right)$ is measured in $\text{s}^{-3}$)
}\label{fig:enstrophy-flux-spectra}
\end{figure}

The development of the direct enstrophy cascade can be traced by means of spectral fluxes of enstrophy, as was done above for the inverse energy cascade.
The classical law is only valid strictly in the inertial range,
where the flux of enstrophy is constant. 
In the presence of bottom friction $\alpha$,
the enstrophy flux $ \Pi_Z $ diminishes from the injection scale towards smaller scales
according to the power law $\Pi_Z (k) \propto k^{-\xi}$ \citep{Valadao:2025,Ventrella:2026}.
Therefore, spectra in flows with friction (drag) behave differently and require a spectral correction $\xi$ to their power laws \citep{Nam:2000,Valadao:2025,Ventrella:2026}:
\begin{equation}
Z\left(k\right)\propto k^{-1-\xi}\left(\ln\left(k/k_{f}\right)\right)^{-1/3}.
\end{equation}
The spectral flux $ \Pi_Z $ of enstrophy is defined similarly to the spectral flux of kinetic energy \citep{Alexakis:2018}:
\begin{equation}
\Pi_{Z}\left(k\right)=\frac{1}{N^{2}}\sum_{k^{\prime}<k}\sum_{k^{\prime}-\frac{1}{2}\leq\left|\boldsymbol{k}\right|<k^{\prime}+\frac{1}{2}}\mathop{\Re\text{e}}\left[\overline{\mathcal{F}\left[\omega\right]\left(\boldsymbol{k}\right)}\cdot\mathcal{F}\left[\boldsymbol{u}\cdot\boldsymbol{\nabla}\omega\right]\left(\boldsymbol{k}\right)\right].
\end{equation}
 The log-log plots of this flux for all the regimes are provided in Fig.~\ref{fig:enstrophy-flux-spectra}.
The slope inside the inertial range corresponds to the spectral correction $\xi$.
Importantly, the enstrophy flux is strictly positive for wavenumbers from the injection scale $k_{f}$ to the hyperdiffusion scale $k\approx k_{h\nu}$, which attests to the transfer of enstrophy from larger to smaller scales, i.~e., to the existence of a direct enstrophy cascade for wavenumbers $k$ in the interval $\left(k_{f}\mathop{..}k_{h\nu}\right)$.

\subsection{Spectral Correction for the Power Law of the Direct Enstrophy Cascade Depending on Friction}\label{subsec:spectral_correction}
% \subsubsection{Spectral correction for the power law of the direct enstrophy cascade depending on friction}

The spectral correction $\xi$ is fitted over the wavenumber segment in Fig.~\ref{fig:enstrophy-flux-spectra} with $k\in\left[2k_{f}\mathop{..}3.8k_{f}\right]$ for all flow regimes (the boundaries of the segment are delineated by dashed vertical lines).
This fitting segment is situated inside the linear portion of the log-log plot of the direct enstrophy cascade spectrum, i.~e., in the inertial range, which is free from forcing-scale and hyperdiffusion-scale contamination.
% The sensitivity of the fitted values to the fitting interval
The fitting-sensitivity errors for the corresponding estimates of $\xi$ are evaluated by varying the
left endpoint of the segment from $1.9k_{f}$ to $3k_{f}$ and the right endpoint of the segment from $3.8k_{f}$ to
$4.4k_{f}$. % \approx2k_{h\nu}/3

\begin{table}[H]
\caption{Spectral correction $ \xi $ to the power law of the direct enstrophy cascade depending on the bottom-friction coefficient $ \alpha $}\label{tab:spectral-correction}
\begin{centering}
\begin{tabular}{c|l|c}
\hline 
$\alpha$, $\unit{\text{s}^{-1}}$ & Regime & $\xi$\tabularnewline
\hline 
\hline 
$0.3$ & laminar--chaotic & $1.7\pm0.1$\tabularnewline
\hline 
$0.1$ & chaotic & $0.60\pm0.03$\tabularnewline
\hline 
$0.01$ & chaotic--vortical & $0.11\pm0.06$\tabularnewline
\hline 
${10}^{-4}$ & vortical & $0.04\pm0.07$\tabularnewline
\hline 
\end{tabular}
\par\end{centering}
\end{table}

The spectral correction estimates are reported in Table~\ref{tab:spectral-correction}.
As $\alpha$ decreases, so too does the correction $\xi$.
For the few values of $ \alpha $ considered here, $ \xi $ is approximately linear in $ \alpha $, with a fitted slope of $ \unit[5.9\pm0.2]{\text{s}} $.
In the limit of $ \alpha\rightarrow0 $, the correction
$ \xi\rightarrow0 $, and the spectrum approaches the classical $k^{-1}$ form:
the enstrophy flux becomes constant, in accordance with the theory \citep{Valadao:2025,Ventrella:2026}.

We recall the exact mathematical identity $ Z\left( k\right) = k^2E\left(k\right) $, valid for incompressible flows in periodic domains \citep{Alexakis:2018} because the Fourier transform of the incompressibility condition is $ \mathcal{F} \left\{ \left( \boldsymbol{\nabla} \cdot \boldsymbol{u} \right) \right\} = ik_x \mathcal{F} \left\{ u \right\} + ik_y \mathcal{F} \left\{ v \right\} =0 $, so
% \begin{equation}
$
\left|\mathcal{F}\left\{ \partial_{x}v-\partial_{y}u\right\} \right|^{2}=\left(k^{2}_{x}+k^{2}_{y}\right)\left(\left|\mathcal{F}\left\{ v\right\} \right|^{2}+\left|\mathcal{F}\left\{ u\right\} \right|^{2}\right).
$
% \end{equation}
 The spectra in our simulations visually coincide for $ k < k_{h\nu} $, a result that provides additional strong evidence that the flow is effectively incompressible.
The maximal relative error $ \left| Z(k) - k^2 E(k) \right| / | Z(k) | $ is at most about $ 4 \% $ in the direct enstrophy cascade with $ k \in \left[k_f \mathop{..} 7k_f\right] $, where $ 7k_f $ is chosen to ensure that the right endpoint is below $ k_{h\nu} $.
Consequently, we observe the dependence $ E\left(k\right) = k^{-2}Z\left(k\right) $ throughout the spectrum, including the direct enstrophy cascade, where $ E\left(k\right) \propto k^{-3-\xi}\left(\ln\left(k/k_{f}\right)\right)^{-1/3} $, so the corresponding spectral corrections extend to the energy spectra.

\section{Conclusions}\label{sec:Conclusion}

This work presents a numerical study of Kolmogorov flow with friction using ILES with implicit sixth-order hyperdiffusion and fixed forcing.
We demonstrate the existence of  statistically stationary laminar, chaotic and vortical regimes in our model, with the bottom-friction coefficient $\alpha$ governing the transition among them.
The effects of compressibility and numerical hyperdiffusion are shown to be negligible in the regimes under consideration.

Linear friction significantly affects both the direct enstrophy and inverse energy cascades.
The direct enstrophy cascade in the presence of bottom friction obeys the following power law with spectral correction $ \xi > 0 $ due to non-constant spectral enstrophy flux: $ Z\left(k\right) \propto k^{-1-\xi}\left(\ln\left(k/k_{f}\right)\right)^{-1/3} $.
We quantify the spectral corrections $ \xi $ as a function of $ \alpha $ using the change in this flux.
In the limit of vanishing friction, the direct enstrophy cascade approaches the shape predicted by the Kraichnan--Batchelor theory with $ \xi \rightarrow 0 $.

Our computational experiments show that the presence of bottom friction also changes the shape of the spectra in the inverse energy cascade.
In a developed vortical regime, bottom friction steepens the spectrum of the inverse cascade of energy, so the inverse cascade appears to obey the law $E\left(k\right)\sim k^{-3}$ observed in domains of finite size.

\section*{Funding}

The present work was supported by the Ministry of Science and Higher Education of the Russian Federation within the framework of the Russian State Assignment No.~124022400174-3 for the KIAM RAS.

\section*{Conflict of Interest}
The authors declare that they have no conflicts of interest.

\sloppy

% \bibliographystyle{unsrt}
% \bibliography{main}
\printbibliography

\fussy
\end{document}